 \documentclass[final,5p,times,twocolumn]{elsarticle}

\usepackage{amssymb}
\usepackage{amsmath}

\usepackage{algorithm}
\usepackage{algorithmic}
\usepackage{graphicx}
\usepackage{caption}
\usepackage{subcaption}
\usepackage{url}

\usepackage[utf8]{inputenc}

\journal{Computer Physics Communications}

\begin{document}

\begin{frontmatter}



\title{Methods for compressible fluid simulation on GPUs using high-order finite differences}


\author[Label1]{Johannes~Pekkil\"a}
\ead{johannes.pekkila@aalto.fi}
\author[Label2]{Miikka~S.~V\"ais\"al\"a}
\ead{miikka.vaisala@helsinki.fi} 
\author[Label3,Label1]{Maarit~J.~K\"apyl\"a}
\author[Label4,Label1,Label3]{Petri~J.~K\"apyl\"a}
\author[Label5,Label1]{Omer~Anjum}

\address[Label1]{ReSoLVE Centre of Excellence, Department of Computer Science, Aalto University, PO Box 15400, FI-00076 Aalto, Finland}
\address[Label2]{Department of Physics, Gustaf H\"allstr\"omin katu 2a, PO Box 64, FI-00014 University of Helsinki, Finland}
\address[Label3]{Max-Planck-Institut f\"ur Sonnensystemforschung, Justus-von-Liebig-Weg 3, D-37077 G\"ottingen, Germany}
\address[Label4]{Leibniz-Institut f\"ur Astrophysik Potsdam, An der Sternwarte 16, D-11482 Potsdam, Germany}
\address[Label5]{Nokia Solutions and Networks, Finland}

\begin{abstract}
We focus on implementing and optimizing a sixth-order finite-difference solver for simulating
compressible fluids on a GPU using third-order Runge-Kutta integration. 
Since \textit{graphics processing units} perform well in data-parallel tasks, this makes them 
an attractive platform for fluid simulation. However, high-order stencil
computation is memory-intensive with respect to both main memory and the caches 
of the GPU. 
We present two approaches for simulating compressible fluids using 55-point
and 19-point stencils. We seek to reduce the requirements for memory
bandwidth and cache size in our methods by using \textit{cache blocking} and decomposing
a latency-bound kernel into several bandwidth-bound kernels. 
Our fastest implementation is bandwidth-bound and integrates $343$ million grid points
per second on a Tesla K40t GPU, achieving a $3.6 \times$ speedup
over a comparable hydrodynamics solver benchmarked on two Intel
Xeon E5-2690v3 processors. Our alternative GPU implementation is latency-bound
and achieves the rate of $168$ million updates per second.\\

\textcopyright 2017. This manuscript version is made available under
the CC-BY-NC-ND 4.0 license, \url{http://creativecommons.org/licenses/by-nc-nd/4.0/} 

Publisher DOI: 10.1016/j.cpc.2017.03.011

\end{abstract}


\end{frontmatter}


\section{Introduction}
\label{introduction}

The number of transistors in a microprocessor has been doubling approximately every
two years and as a result, the performance of supercomputers measured in 
\textit{floating-point operations per second} (FLOPS) has been following a similar increase. 
However, since increasing the clock frequencies of microprocessors to gain better performance 
is no longer feasible because of power constraints, this has lead to a change in their architectures
from single-core to multi-core. 

\par

While modern \textit{central processing units} (CPUs) utilize more cores and wider
SIMD units, they are designed to perform well in general tasks where low memory
access latency is important. 
On the other hand, \textit{graphics processing units} (GPUs) are specialized in solving
data-parallel problems found in real-time computer graphics and as a result, house more
parallel thread processors and use higher-bandwidth memory than CPUs.
With the introduction of
general-purpose programming frameworks, such as \textit{OpenCL} and
\textit{CUDA}, GPUs can now also be programmed to do general purpose tasks
using a C-like language instead of using a graphics \textit{application-programming interface} (API),
such as \textit{OpenGL}. In addition, APIs such as \textit{OpenACC} can be used
to convert existing CPU programs to work on a GPU. For these reasons, GPUs offer an
attractive platform for physical simulations which can be solved in a data-parallel fashion.

\par

In this work we concentrate on investigating sixth-order central
  finite-difference scheme implementations on GPUs, suitable for multiphysics
applications.
The justification for the use of central differences with 
explicit time stepping, a configuration which is not ideal concerning
its stability properties, comes from the fact that, even though some 
amount of diffusion is required for stability,
they provide very good accuracy and are easy to implement (see,
e.g. \cite{brandenburg}). 
In addition, the various types of boundary conditions and grid geometries needed
in multiphysics codes such as the Pencil Code\footnotemark[1] are easy to implement
with central schemes.
Moreover, the problem has the potential to exhibit strong scaling with
the number of parallel cores in the optimal case.
\footnotetext[1] {http://github.com/pencil-code} 

\par

There are astrophysical hydro- and
magnetohydrodynamic solvers already modified to take advantage of
accelerator platforms (i.e. \cite{ChollaCode}, \cite{ENZOcode},
\cite{SmaugCode}), that most often use low-order discretization.
As an example of a higher-order scheme for cosmological hydrodynamics,
we refer to \cite{Cosmol_5th}.
We also note that more theoretical than application-driven work on
investigating higher-order stencils on GPU architecture exists in the
literature, see e.g. \cite{Zumbusch2013}.  
There are many scientific problems, such as modeling
hydromagnetic dynamos, where long integration times are required,
either to reach a saturated state (see e.g.\ \cite{gent2013}), or to 
exhibit non-stationary phenomena and secular
trends (see e.g.\ \cite{kapyla2015}). Therefore, it is highly desirable
to find efficient ways to accelerate the methods, GPUs offering an
ideal framework.
The accelerated codes typically employ lower-order conservative schemes,
in which case the halo region to be communicated to compute the
differences is small, and does not pose the main challenge for the
GPU implementation. High-order schemes of similar type as presented
here exist for two-dimensional hydrodynamics (e.g.\ \cite{CMB12});
in this paper, we deal with
a 3D implementation of a higher-order finite-difference solver.
Such schemes are much less diffusive and they are more suitable
for accurate modeling of turbulence, which
is, on the other hand, crucial for e.g. investigating various types of
instabilities in astrophysical settings. One mundane example, which is 
the solar dynamo, is responsible for all the
activity phenomena on the Sun, driving the space weather and
climate that affect life on Earth \cite{2014Charbonneau}.
The accurate modeling of turbulence is also important in understanding such 
phenomena as the structure of interstellar medium \cite{2004ElmegreenScalo} 
and star formation \cite{2007McKeeOstriker}. 

\par

We make the following contributions in this work. First, we describe, implement and optimize
two novel methods for simulating compressible fluids on GPUs using sixth-order finite
differences and 19- and 55-point stencils.
The current implementation is for simulations of isothermal fluid
  turbulence.
  The bigger picture is that it uses the same core methods as the
  Pencil Code. Thus the current code
development works as a pilot project in the conversion of the
Pencil Code to use GPUs.

Our implementations perform $1.7 \times$ and $3.6 \times$ faster 
than a state-of-the-art finite difference solver, Pencil Code, used for scientific computation on HPC-clusters.
Second, we present an optimization technique called \textit{kernel decomposition}, which can be 
used to improve the performance of latency-bound kernels. Currently our
code, called \textit{Astaroth}, supports isothermal compressive hydrodynamics, but it will be expanded in the future
to include more complex physics, in the end supporting the full equations of 
magnetohydrodynamics (MHD). 

\par

In this paper, we present the physical motivation (Sect. \ref{sec:problem_specification})
behind our implementations, and the technical justification and background (Sect. \ref{sec:gpu_architecture}).
The details of our implementations and the Astaroth code are presented in Sect. \ref{sec:implementations}.
In Sect. \ref{sec:results} we present the performance of our GPU implementations 
and compare the results with physical test cases in Sect. \ref{sec:testcases}. Finally, in Sect. \ref{sec:discussion},
we discuss our results and conclude the paper.

\section{Problem specification}
\label{sec:problem_specification}

Here we describe the basic equations and numerical methods featured in the 
current implementation of the Astaroth code, which were also used in testing 
the different optimizations. For simplicity, the code is limited to the domain 
of hydrodynamics. We consider the fluid to be isothermal and compressible, 
and we include the full formulation of viscosity to the momentum equation. 
This allows for testing our methods with reasonable enough physics while 
avoiding overt complexity during the development of the methods.

\label{sec:equations}

\subsection{Governing equations}

In a compressible system, the conservation of mass can be
expressed as the rate equation for density $\rho$, called the \textit{continuity equation}:
\begin{small}
\begin{equation}
\frac{D \ln\rho}{D t} = - \nabla \cdot \mathbf{u}.
\label{eq:continuity}
\end{equation}
\end{small}
Here $D / Dt$ is the convective derivative 
$\partial / \partial t + (\mathbf{u} \cdot \nabla )$ and density is 
expressed in
logarithmic form $\ln\rho$ and $\mathbf{u}$ is a 
three-dimensional velocity vector. 
The logarithmic form of density helps to avoid numerical errors that can occur
with large stratifications or erroneously negative values of density.

Momentum conservation in a viscous fluid is modelled by a rate equation commonly known as the
\textit{Navier-Stokes equation}. In the case of 
isothermal viscous hydrodynamics featured in the Astaroth code it is given by:
\begin{small}
\begin{equation}
\frac{D \mathbf{u}}{D t} = -c_s^2 \nabla \ln\rho + \mathbf{f} +\nu \bigg( \nabla^2\mathbf{u} + 
\frac{1}{3}\nabla(\nabla \cdot \mathbf{u}) + 2 \mathbf{S} \cdot \nabla \ln \rho \bigg),
\label{eq:navierstokes}
\end{equation}
\end{small}
where $-c_s^2 \nabla \ln\rho$ is the pressure term 
in the isothermal case where the pressure is given by $p=\rho c_s^2$
with $c_s$ being the constant sound speed,
$\mathbf{f}$ is an external body force, such as an external gravity field or a
forcing function (see Sect.~\ref{sec:testcases}),
$\nu$ is the kinematic viscosity coefficient, which is assumed constant, and $\mathbf{S}$ is the traceless
rate-of-strain tensor: 
\begin{equation}
S_{ij} = \frac{1}{2} \bigg( \frac{\partial u_i}{\partial x_j} + 
\frac{\partial u_j}{\partial x_i} - \frac{2}{3}\delta_{ij} \nabla \cdot \mathbf{u} \bigg).
\label{eq:tensor}
\end{equation}

\subsection{Non-dimensional units and system parameters}

While solving the equations, we assume that the variables are described in 
dimensionless manner. Depending on the nature of the actual physical question, 
the result can be scaled during the data analysis phase into relevant physical 
dimensions. In this way, we can avoid using numerically unsound parameter 
values. 

The dimensionless physical units are defined as
\begin{small}
\begin{equation}
\label{eq:dimensionless}
\begin{split}
\rho = \frac{\rho_\mathrm{phys}}{M_\mathrm{u}/L_\mathrm{u}^3}, & 
\quad \mathbf{u} = \frac{\mathbf{u}_\mathrm{phys}}{L_\mathrm{u}/T_\mathrm{u}}, 
\quad \nu = \frac{\nu_\mathrm{phys}}{L_\mathrm{u}^2/T_\mathrm{u}}, \\
c_s = \frac{c_{s\mathrm{,phys}}}{L_\mathrm{u}/T_\mathrm{u}}, &
\quad k = \frac{k_\mathrm{phys}}{2\pi / L_\mathrm{u}} 
\end{split}
\end{equation}
\end{small}
where $L_\mathrm{u}$, $M_\mathrm{u}$ and $T_\mathrm{u}$ denote chosen 
unit scaling of length, mass and time respectively.

In addition, we use an important dimensionless measure, the Reynolds number,
\begin{small}
\begin{equation}
  \mathrm{Re} = \frac{u L}{\nu}
\label{eq:reynolds}
\end{equation}
\end{small}
where $u$ and $L$ represent characteristic velocities and length scales in the system.

\subsection{The finite difference method}
\label{sec:finite_difference}

We discretize the fluid volume $(L_x,L_y,L_z)$ onto an equidistant
grid of $(N_x, N_y,N_z)$ points, the distance between neighbouring grid
points in each dimension being $\delta_i=L_i/N_i$. 
With these, the approximation of the first derivative of
function $f$ on the grid point $i$ with respect to the $x$-direction when
using sixth-order central differences can be written as follows
\cite[pp.~5--7]{brandenburg}. 

\begin{align}
\begin{split}
\frac{\partial}{\partial x} f_{i} = \frac{1}{60\delta_x} 
						\biggl(  	&- f_{i-3} + 9 f_{i-2} - 45 f_{i-1}\\
								&+ 45 f_{i+1} - 9 f_{i+2} + f_{i+3} 
						\biggr) + O(\delta_x^6) \ .
\end{split}
\label{eq:first_derivatives}
\end{align}

\noindent The second derivatives can be approximated with

\begin{align}
\begin{split}
\frac{\partial^{2}}{\partial x^{2}} f_{i} = \frac{1}{180\delta_{x}} 
						\biggl(&2f_{i-3} - 27f_{i-2} + 270f_{i-1} - 490f_{i}\\
						             &+270f_{i+1} - 27f_{i+2} + 2f_{i+3}
						\biggr) + O(\delta_{x}^{6}) \ .
\end{split}
\label{eq:second_derivatives}
\end{align}

\noindent Additionally, mixed derivatives with respect to any two arbitrary
directions can be approximated by using the following bidiagonal scheme \cite{pencil_code}, here
with respect to $x$- and $y$-directions.

\begin{align}
\begin{split}
\frac{\partial^{2}}{\partial x \partial y} f_{i,j} = \frac{1}{720\delta_{x}\delta_{y}}
			&\biggl[ 270( f_{i+1, j+1} - f_{i-1, j+1} + f_{i-1, j-1} - f_{i+1, j-1} )\\
			&- 27 ( f_{i+2, j+2} - f_{i-2, j+2} + f_{i-2, j-2} - f_{i+2, j-2} )\\
			&+ 2  ( f_{i+3, j+3} - f_{i-3, j+3} + f_{i-3, j-3} - f_{i+3, j-3} )
			\biggr]\\
                        &+ O(\delta_{x}^{6}, \delta_{y}^{6})\ .
\end{split}
\label{eq:mixed_derivatives}
\end{align}

\noindent Finally, we can approximate second and mixed derivatives using Eq. \eqref{eq:first_derivatives}
as follows

\begin{equation}
\begin{aligned}
\label{eq:19-point_mixed_derivatives}
\frac{\partial}{\partial x}(\frac{\partial}{\partial y} f_{i,j}) = 
			\frac{1}{60\delta_{x}}
			\biggl[
			&-\biggl(\frac{\partial}{\partial y}f_{i-3, j}\biggr) 
			+ 9\biggl(\frac{\partial}{\partial y}f_{i-2, j}\biggr) 
			- 45\biggl(\frac{\partial}{\partial y}f_{i-1, j}\biggr)\\
			&+ 45\biggl(\frac{\partial}{\partial y}f_{i+1, j}\biggr) 
			- 9\biggl(\frac{\partial}{\partial y}f_{i+2, j}\biggr) 
			+ \biggl(\frac{\partial}{\partial y}f_{i+3, j}\biggr) 
			 \biggr]\\
			&+ O(\delta_{x}^{6}) \ .
\end{aligned}
\end{equation}

\begin{figure}
\centering
\includegraphics[width=.4\linewidth]{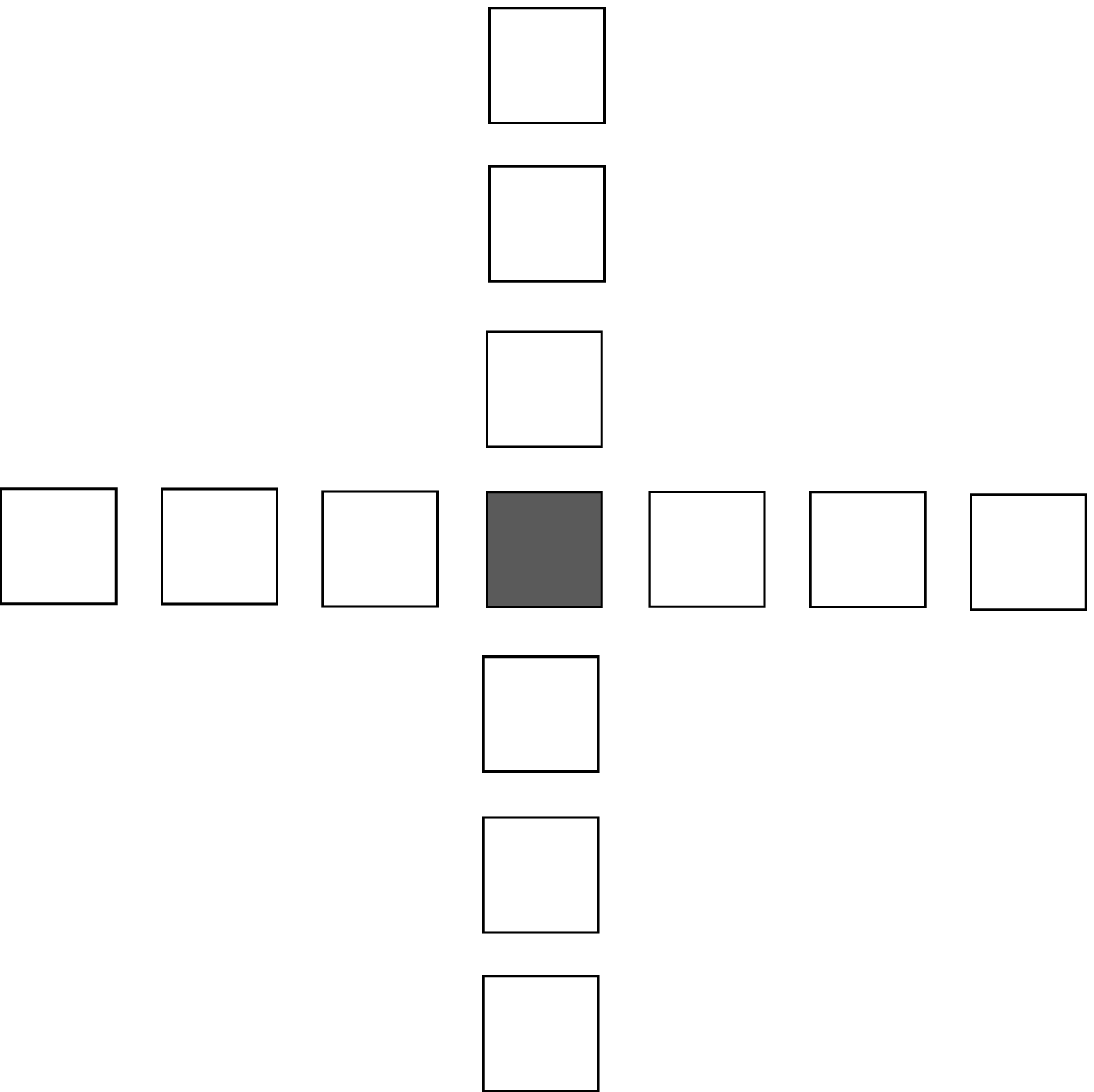}
\hspace{10pt}
\includegraphics[width=.4\linewidth]{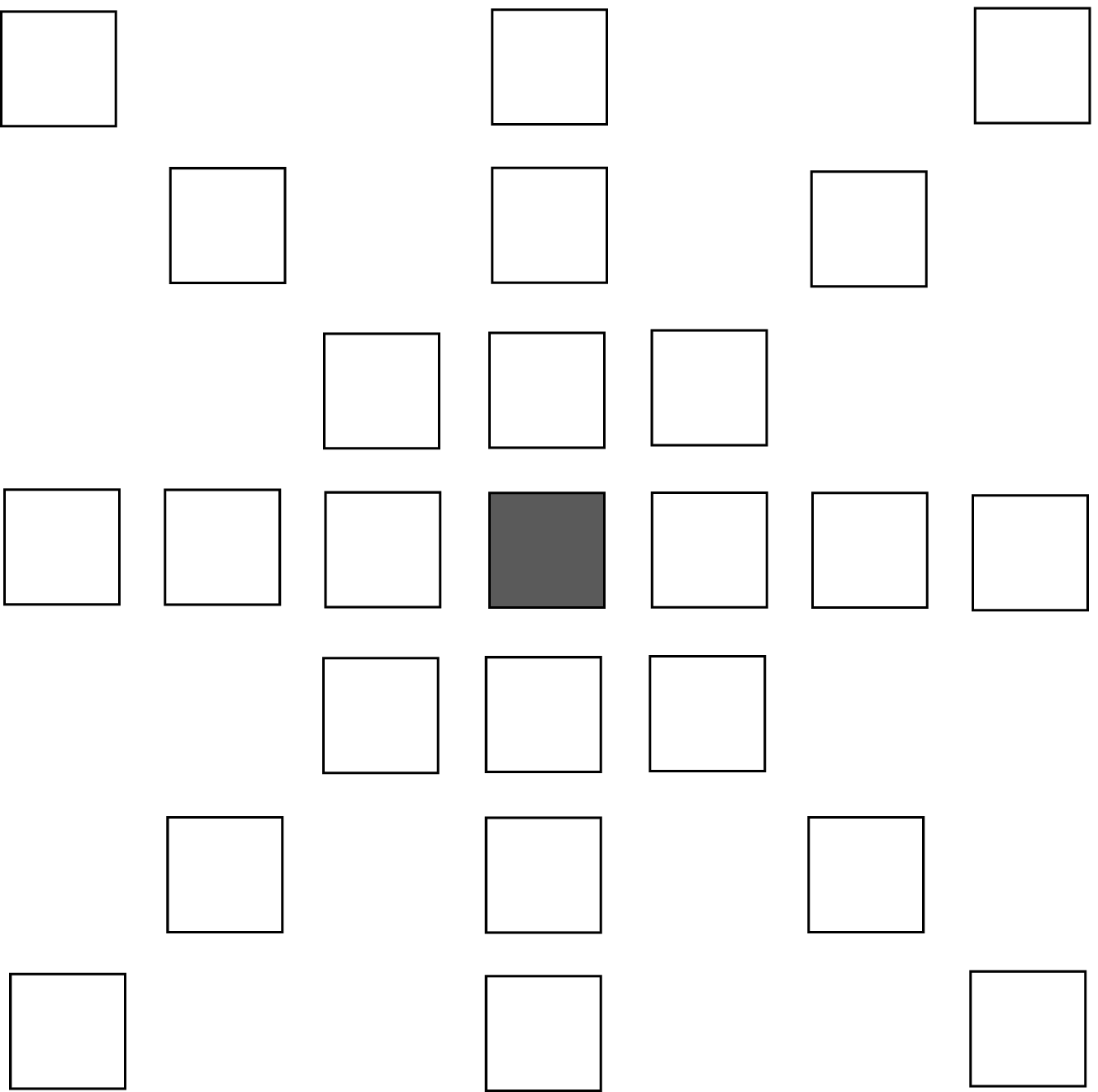}
\caption{Two-dimensional cuts of the stencils used in the functions, which approximate derivatives using sixth-order finite 
differences. A 19-point stencil (\textit{left}) is used in a function, which solves only first- and second-order 
derivatives, whereas a 55-point stencil (\textit{right}) is used in a function, which computes also mixed derivatives 
using the bidiagonal scheme. The white and dark cells are input and output points, respectively. For both
stencils, $R=3$.}
\label{fig:stencils}
\end{figure}

Computing the derivatives of a grid point requires information from its neighboring 
grid points. This data access pattern is called a $k$-point stencil, where data from $k$ input points is 
read in order to update the output point. We use $R$ to denote the \textit{radius} of the stencil, 
which is the Chebychev distance from the output point to the farthest input point of the stencil. For 
central differences, the relation between the radius of the stencil $R$ and the order of the finite 
difference method $n$ is thus $n = 2R$. Additionally, a function solving only the first- and 
second-order derivatives in $d \geq 1$ dimensions using an $n$th-order central finite difference method 
uses a $(dn + 1)$-point stencil to update a grid point, whereas a function, which computes also the 
mixed derivatives requires a $(dn + 2\binom{d}{2}n + 1)$-point stencil for $d \geq 2$. The stencils used in such 
functions are shown in Fig. \ref{fig:stencils}.

\subsection{Runge-Kutta integration}
\label{sec:2N_RK_scheme} 

Our implementations are based on an explicit third-order Runge-Kutta formula, 
which is written as a $2N$-storage scheme \cite[p.~9]{brandenburg}. In this
approach, only one set of intermediate values have to be memory-resident during
integration.

\par

Let $\mathbf{u}$ be a vector
field the integration is performed on and $\widetilde{\mathbf{u}}$ be the field containing
the intermediate values. Additionally, let 
$\mathbf{u}^{(s)}$ be the value of $\mathbf{u}$ during integration substep $s$. Finally, 
$\alpha^{(s)}$ and $\beta^{(s)}$ are coefficients whose values depend on the chosen 2N-RK3 scheme
and $\delta_{t}$ is the length of the time step.

\par

We can now write the 2N-RK scheme as

\begin{align}
\label{eq:2N-RK_scheme}
\widetilde{\mathbf{u}}^{(s+1)} = \alpha^{(s)} \widetilde{\mathbf{u}}^{(s)} +
\frac{\partial}{\partial t} \mathbf{u}^{(s)} \delta_{t} \qquad \mathbf{u}^{(s+1)} = \mathbf{u}^{(s)} + \beta^{(s)} \widetilde{\mathbf{u}}^{(s+1)} \ .
\end{align}

The pseudocode for a na\"{\i}ve integration with this scheme is shown in Algorithm \ref{alg:2N_runge-kutta}.
Here
$\widetilde{\rho}_{i}$ and $\widetilde{\mathbf{u}}_{i}$ are the
intermediate results for density and velocity of a grid point at index $i$.
Handling of the out-of-bound indices depends on the chosen boundary condition scheme.
For the first step, $\alpha^{(1)}$ must be set to $0$.

\begin{algorithm}
\caption{Third-order 2N-storage Runge-Kutta integration}
\begin{algorithmic}
\REQUIRE Integer $i$ belongs to the set of indices in the computational domain $D$.
\FOR{integration substep s = 1 \TO 3 }
    \STATE{Compute boundary conditions for $\rho$ and $\mathbf{u}$}
	\FORALL {$i \in D$ \textbf{in parallel}} 
        \STATE {$\widetilde{\rho}_{i} \leftarrow \alpha^{(s)}\widetilde{\rho}_{i} + \delta t\frac{\partial}{\partial t} \rho_{i}$}
        \STATE {$\widetilde{\mathbf{u}}_{i} \leftarrow \alpha^{(s)}\widetilde{\mathbf{u}}_{i} + \delta t\frac{\partial}{\partial t} \mathbf{u}_{i}$}
	\ENDFOR 
	\FORALL {$i \in D$ \textbf{in parallel}} 
        \STATE {$\rho_{i} \leftarrow \rho_{i} + \beta^{(s)}\widetilde{\rho}_{i}$}
        \STATE {$\mathbf{u}_{i} \leftarrow \mathbf{u}_{i} + \beta^{(s)}\widetilde{\mathbf{u}}_{i}$}
	\ENDFOR 
\ENDFOR
\end{algorithmic}
\label{alg:2N_runge-kutta}
\end{algorithm}

\section{GPU architecture}
\label{sec:gpu_architecture}

In this section, we review GPU architecture using NVIDIA's CUDA as the framework of choice and
discuss the challenges of high-order stencil computation on GPUs. While we use terminology associated
with CUDA, the ideas represented here are also analogous with those found in OpenCL and computer architecture 
in general. We denote the alternative terminology in the footnotes. Throughout this work, we use NVIDIA's 
compute capability 3.5 GPUs as the baseline architecture. 

\par

Graphics processing units operate in a multi-threaded SIMD fashion and are designed to perform well in
data-parallel tasks. In order to maximize throughput in these 
types of tasks, GPUs employ a large number of parallel \textit{thread processors}\footnotemark[1] and use 
specialized GDDR SGRAM to increase memory bandwidth with the cost of increased access latency. Modern 
GPUs also employ small L1 and L2 caches to reduce pressure to the on-device memory. See
Table \ref{tab:hardware_limits} for the detailed specifications of a Tesla K40t accelerator card used in this work.
In order to hide pipeline and memory access latencies, GPUs rely mainly on multithreading a large number of threads on their 
processors in a fine-grained fashion. Alternatively, in certain problems \textit{instruction-level parallelism} (ILP) can be 
used for the same latency-hiding effect \cite{volkov2008}.

\par

Modern NVIDIA GPUs consist of \textit{Streaming Multiprocessors}\footnotemark[2] (SMX), which execute 
\textit{warps}\footnotemark[3] of \textit{CUDA threads}\footnotemark[4]. In current NVIDIA architectures, a 
warp is composed of $32$ threads. The threads of a warp are executed in lockstep on the thread processors of 
an SMX. Finally, sets of warps form \textit{thread blocks}\footnotemark[5], which are distributed among
SMXs by a thread block scheduler. For a more detailed description of the architecture of GPUs, we refer the 
reader to \cite{computer_organization_book} and \cite{computer_architecture_book}.

\footnotetext[1]{Analogous terms: Processing Element, SIMD lane, CUDA core.}
\footnotetext[2]{Analogous terms: Compute Unit, multi-threaded SIMD processor.}
\footnotetext[3]{Analogous terms: Wavefront, SIMD thread.}
\footnotetext[4]{Analogous terms: Work item, instruction stream of a SIMD thread.}
\footnotetext[5]{Analogous terms: Work group.}

\begin{table}[h]
\caption{Tesla K40t specifications rounded to two decimal places.}
\begin{center}
    {\small
    \begin{tabular}{ | l | l |}	
    \hline
     & Tesla K40t\\
	\hline
	GPU chip & GK110BGL\\
	\hline
	Compute capability & 3.5\\
	\hline
	GPU memory (GDDR5 SGRAM) & 12288 MiB\\
	\hline
	Memory bus width & 384 bits\\
	\hline
	Peak memory clock rate & 3004 MHz\\ 
	\hline
	Theoretical memory bandwidth & 268.58 GiB/s\\
	\hline
	Number of SMX processors & 15\\
	\hline
	Max 32-bit registers per SIMD processor & 65 536\\
	\hline
	Max shared memory per thread block & 49 152 bytes\\
	\hline
	L2 cache size & 1.50 MiB\\
	\hline
    \end{tabular}
    }
\end{center}
\label{tab:hardware_limits}
\end{table}

\subsection{Related work and optimization techniques}

Optimization of GPU programs is often non-trivial and requires careful tuning to
attain the highest throughput. Techniques for optimizing low-order stencil computation
have been studied extensively in literature, e.g. by \cite{mikki2009} \cite
{nguyen2010} \cite{zhang2012} \cite{giles2014} \cite{vizitiu2014}, while work on higher-order stencil computation is 
more limited \cite{cruz2014} \cite{cygert2014} and focuses on stencils, where only the axis-aligned elements are used 
during an update. To our knowledge, no previous work has been published on optimizing 55-point stencil computation 
using stencil elements of size $16$ bytes and larger.

\par

Memory bandwidth is a common bottleneck in stencil computation because of the low bandwidth-to-compute ratio in 
current microprocessor architectures. Previous work has suggested several techniques for reducing bandwidth 
requirements, notably spatial cache blocking techniques which aim to reduce memory fetches by reusing as much of 
the data as possible before evicting it out of caches. We also adopted this idea in our implementations we discuss in 
detail in Sect. \ref{sec:55-point_integration_method} and \ref{sec:19-point_integration_method}.

\par

However, with higher-order stencils more data is required to be cache-resident in order to improve performance with 
cache blocking. Using large amounts of shared memory in a kernel reduces its occupancy, which in turn results in decreased 
ability to hide latencies if increasing ILP in the kernel is not possible. In current GPU architectures, the cache size is 
insufficient for housing large three-dimensional blocks of $16$-byte-sized grid points.

\section{GPU Implementations}
\label{sec:implementations}

Here we give an overview of our GPU implementations and the optimizations performed upon both 
implementations. First, we discretize the simulation domain into a grid as described in Sect. \ref
{sec:finite_difference}. The grid consists of a \textit{computational domain}, which is surrounded by a \textit
{ghost zone}. We update the grid points in the computational domain using third-order Runge-Kutta integration, while the ghost 
zone is used to simplify integration near the boundaries. After each integration step, a number of grid points 
are copied from the computational domain into the ghost zone according to the boundary conditions. With this
approach, we do not have to compute the boundary conditions within the integration kernel and can thus update
all grid points in the computational domain using the same code. We use periodic boundary conditions throughout
this work.

\par

We store density $\rho$, velocity $\mathbf{u}$ and the intermediate result arrays $\widetilde{\rho}$ and $
\widetilde{\mathbf{u}}$ into global memory as a structure of arrays. Arrays $\rho$ and $\mathbf{u}$ 
include the ghost zone and are padded manually, such that the first grid point of the computational domain 
is stored in a memory address which is a multiple of 128 bytes. By padding, we seek to reduce the number 
of memory transactions required to update the grid, which is discussed more in detail in Sect.
\ref{sec:results}.
In order to avoid a race condition during integration, we allocate memory for arrays
$\rho$, $\mathbf{u}$, $\rho^{\text{out}}$ and $\mathbf{u}^{\text{out}}$, such that separate arrays are used
for reading and writing. Arrays $\rho$ and $\mathbf{u}$ are passed to the integration kernels using the
\verb!const __restrict__! intrinsic, which enables these arrays to be read through the read-only texture cache.
Additionally, we store all constants discussed in Sect. \ref{sec:problem_specification} into constant memory.
Finally, in both our integration kernels, we compute each derivative only once and store each value,
which is used more than once, into either shared memory or registers.

\subsection{55-point integration method}
\label{sec:55-point_integration_method}

We solve the problem on a GPU using
a modified version of Algorithm \ref{alg:2N_runge-kutta}. The 
difference is, that we use separate arrays for reading and writing and thus can update the computational domain
in one pass over the grid points. At the end of each integration substep, the arrays are swapped efficiently
using pointers.

\par

In our modified integration algorithm, we update a grid point by solving the continuity and Navier-Stokes
equations, Eq. \eqref{eq:continuity} and \eqref{eq:navierstokes}, within a single kernel.
As we  solve the derivatives in these equations with the finite-difference equations \eqref{eq:first_derivatives}, 
\eqref{eq:second_derivatives} and \eqref{eq:mixed_derivatives}, the integration kernel requires a 55-point stencil in order to update a 
grid point. Therefore we call this approach the \textit{55-point method}. The pseudocode for the 55-point method is shown in Algorithm \ref{alg:55-point_method} using the notation 
introduced in Sections \ref{sec:2N_RK_scheme} and \ref{sec:implementations}.

\begin{algorithm}
\caption{Integration with the 55-point method}
\begin{algorithmic}
\REQUIRE Integer $i$ belongs to the set of indices in the computational domain $D$.
\FOR{integration step s = 1 \TO 3 }
	\STATE{Compute boundary conditions for $\rho$ and $\mathbf{u}$}
	\FORALL {$i \in D$ \textbf{in parallel}} 
	\STATE {$\widetilde{\rho}_{i} \leftarrow \alpha^{(s)}\widetilde{\rho}_{i} + \delta t\frac{\partial}{\partial t} \rho_{i}$}
	\STATE {$\widetilde{\mathbf{u}}_{i} \leftarrow \alpha^{(s)}\widetilde{\mathbf{u}}_{i} + \delta t\frac{\partial}{\partial t} \mathbf{u}_{i}$}
	\STATE {$\rho_{i}^{\text{out}} \leftarrow \rho_{i}+ \beta^{(s)}\widetilde{\rho}_{i}$}
	\STATE {$\mathbf{u}_{i}^{\text{out}} \leftarrow \mathbf{u}_{i} + \beta^{(s)}\widetilde{\mathbf{u}}_{i}$}
	\ENDFOR 
	 
	 \STATE{$\rho \leftarrow \rho^{\text{out}}$}
	 \STATE{$\mathbf{u} \leftarrow \mathbf{u}^{\text{out}}$}
\ENDFOR
\end{algorithmic}
\label{alg:55-point_method}
\end{algorithm}

\par

The algorithm is implemented in CUDA as follows. 
Let $\tau_x$, $\tau_y$ and $\tau_z$ be the dimensions of a thread block and $R$ the radius
of the stencil as defined in Sect. \ref{sec:finite_difference}. We perform the integration by decomposing the 
computational domain into $\tau_x \times \tau_y \times \tau_z$-sized blocks, where each grid point is 
updated by a CUDA thread. Since the stencils used to update nearby grid points overlap, we can reduce global 
memory fetches by fetching the data used by the threads in a thread block into shared memory.

\par

\begin{figure}
\centering
\includegraphics[width=0.8\linewidth]{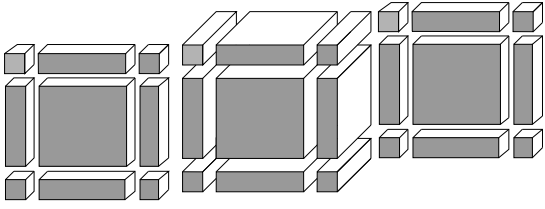}
\caption{A dissected shared memory block used by the 55-point method. The halo areas in the front and 
back of the block have been moved to left and right for clarity.}
\label{fig:55p_smem_block}
\end{figure}

The block of grid points stored into shared memory per a thread block is shown in Fig. \ref
{fig:55p_smem_block}. We call the area surrounding the shared memory block a \textit{halo} in order to 
distinguish it from the ghost zones discussed in Sect. \ref{sec:implementations}. Unlike ghost zones, the 
boundary conditions are not applied to the halo and the grid points in the halo are solely used for updating the 
grid points near the boundaries of the thread block. For simplicity, we fetch a total of $(\tau_x+2R) \times 
(\tau_y+2R) \times (\tau_z+2R)$ grid points into shared memory per a thread block, even when some of the 
grid points are used only by a single thread or none at all. This approach avoids branching in the integration 
kernel, as all threads follow the same execution path with the cost of additional memory fetches. We discuss these
redundant memory fetches more in detail in Sect. \ref{sec:discussion}.

\par

In order to reduce the number of memory transactions from global memory, we adopted the idea of a rolling 
cache. In this approach, part of the data in shared memory is reused for updating multiple grid points 
along the $z$-axis. We implemented cache blocking for Alg. \ref{alg:55-point_method} as follows.

\noindent\textbf{Initial step:}\\
(a) Assign a block of grid points in the decomposed computational domain to a CUDA thread block.\\
(b) Fetch the data required for updating this block of grid points from global memory
into shared memory\footnote{$(\tau_x+2R) \times (\tau_y+2R) \times (\tau_z+2R)$ grid points}.\\ 
(c) Update the block of grid points using the data stored in shared memory.\\

\noindent\textbf{Subsequent steps:} While a thread block has updated less than $E_{z}$ blocks of grid points, do the following:\\
(d) Assign the next block of grid points in the $z$-axis to the thread block.\\
(e) Since the halos of nearby blocks overlap, part of the
data obtained in the previous step can also be used to update the current block of grid
points\footnote{$(\tau_x + 2R) \times (\tau_y + 2R) \times 2R$ grid points}.
Hold this data in shared memory. Load rest of the required data from global memory
into shared memory\footnote{$(\tau_x + 2R) \times (\tau_y + 2R) \times \tau_z$ grid points}.\\
(f) Update the assigned block of grid points using the data in shared memory.

\par

In our implementation of the rolling cache, we avoid copying data around shared memory by using a counter indicating the current 
mid-point in the shared memory block. After updating a grid point, we increment this counter by $\tau_z$ in 
modulo $\tau_z+2R$. During differentiation, any out-of-bound indices encountered when accessing shared memory are also wrapped around modulo $\tau_z+2R$.

\par

Additionally, let $N_{x}$, $N_{y}$ and $N_{z}$ be the dimensions of the computational domain and 
$E_{z}$ the number of grid points updated by a CUDA thread. The total number of thread blocks required to 
update the grid is thus

\begin{equation}
\label{eq:thread_blocks_per_grid}
\gamma = \frac{N_{x}}{\tau_{x}} \cdot \frac{N_{y}}{\tau_{y}} \cdot \frac{N_{z}}{\tau_{z} E_{z}} \ ,
\end{equation}

and the number of grid points fetched from global memory is

\begin{equation}
\Gamma_{55p} = (\tau_{x} + 2R) \times (\tau_{y} + 2R) \times (\tau_{z}E_{z} + 2R) \ .
\end{equation}

With the 55-point method, we perform the following number of read-writes $RW_{55p}$ to global memory when $N_{xyz} = N_{x}N_{y}N_{z}$. 
Here we read from four arrays which require the halo ($\rho$ and $\mathbf{u}$) and four intermediate value arrays ($\widetilde{\rho}$ and $\widetilde{\mathbf{u}}$), and write the result back to eight arrays ($\rho^{\text{out}}$, $\mathbf{u}^{\text{out}}$, $\widetilde{\rho}$ and $\widetilde{\mathbf{u}}$).

\begin{equation}
\label{eq:55p_rw}
RW_{55p} = \underbrace{4 \Gamma_{55p} \gamma + 4 N_{xyz}}_{\text{Reads}} + \underbrace{8 N_{xyz}}_{\text{Writes}} \ .
\end{equation} 

 \par

However, the main problem with the 55-point method is low occupancy caused by the large amount
of shared memory required by a thread block in order to benefit from cache blocking. This is the case 
also for small thread blocks. For example, when using thread blocks of size $\tau_x = \tau_y = \tau_z = 8$, $R=3$ and 
storing $16$ bytes of information per grid point, then a thread block requires $43
\ 904$ bytes of the $49\ 152$ bytes available shared memory on a Tesla K40t GPU. Therefore only one 
thread block can run on the GPU at a time. This is not enough to hide the latencies in our integration kernel, 
which becomes latency-bound. Moreover, instruction-level parallelism cannot be used extensively to hide the 
latencies in our approach, since the data in shared memory can be updated only after the threads of a thread 
block have updated their currently assigned grid points.

\subsection{19-point integration method}
\label{sec:19-point_integration_method}

To alleviate the problem with high shared memory usage in our 55-point method, we represent an alternative 
integration method which uses an axis-aligned 19-point stencil to update a grid point. This is achieved by 
computing the gradient of divergence in Eq. \ref{eq:navierstokes} in two passes over the grid. The benefit of this 
approach is, that stencil computation on GPU with axis-aligned stencils is extensively studied and efficient cache 
blocking methods for such stencils are well known \cite{mikki2009} \cite{nguyen2010} \cite{zhang2012}. However, 
the disadvantages of this approach compared with our 55-point method are three-fold: first, we have to perform 
more floating-point arithmetic in order to update a grid point, which introduces a slight error. We show in Sect. \ref
{sec:testcases} that this error is negligible.
Second, as the grid is updated in two
steps, more memory transactions are required 
to complete a single integration step. Third, when solving the system with multiple 
GPUs, part of the divergence field has to be communicated between the nodes. 

\par

The 19-point method works as follows. First, we reformulate the Navier-Stokes equation in such form, that mixed 
derivatives do not have to be solved in order to compute the gradient of divergence. This is achieved by dividing a 
substep of a full integration step into two passes, where the divergence field is solved during the first pass and 
during the second pass, the 
gradient of divergence is solved using the precomputed divergence field. The complete 
reformulation of the Navier-Stokes equation is shown in Appendix A. Using the notation from Sect. \ref
{sec:2N_RK_scheme} and $\widetilde{\mathbf{u}}_{partial}^{(s)}$ and $\mathbf{u}_{partial}^{(s)}$ to denote the 
partially computed Navier-Stokes equation, we can write the calculations done during the first pass as follows. 
Here $s$ denotes some substep of a full integration step. For RK3, $s \in \{1, 2, 3\}$.

\begin{align}
\nabla \cdot \mathbf{u}^{(s)}
&= \frac{\partial}{\partial x} u + \frac{\partial}{\partial y} v + \frac{\partial}{\partial z} w
\\
\widetilde{\mathbf{u}}_{partial}^{(s+1)} 
&= \alpha^{(s)} \widetilde{\mathbf{u}}^{(s)} + \delta_{t} \biggl[ -({\mathbf{u}}^{(s)}\cdot\nabla){\mathbf{u}}^{(s)}-c_s^2 \nabla \ln\rho^{(s)} 
\nonumber \\ &+ \nu \bigg( \nabla^2\mathbf{u}^{(s)} + 2 \mathbf{S} \cdot \nabla \ln \rho^{(s)} \biggr)  \biggr] 
\\
\mathbf{u}_{partial}^{(s+1)}
&= \mathbf{u}^{(s)} + \beta^{(s)} \widetilde{\mathbf{u}}_{partial}^{(s+1)} \ ,
\end{align}
where $\mathbf{u}=(u,v,w)$.

Then, with the second pass we complete the integration step by computing 
$\widetilde{\mathbf{u}}^{(s+1)}$ and $\mathbf{u}^{(s+1)}$ using the previously computed
divergence field and the partial results.

\begin{align}
\widetilde{\mathbf{u}}^{(s+1)} 
&= \widetilde{\mathbf{u}}_{partial}^{(s+1)} + \delta_{t} \frac{\nu}{3}\nabla(\nabla \cdot \mathbf{u}^{(s)})
\\
\mathbf{u}^{(s+1)} 
&= \mathbf{u}_{partial}^{(s+1)} + \beta^{(s)} \delta_{t} \frac{\nu}{3}\nabla(\nabla \cdot \mathbf{u}^{(s)}) \ .
\end{align}
The pseudocode for this approach is shown in Algorithm \ref{alg:19-point_method}.
\begin{algorithm}
\caption{Integration with the 19-point method}
\begin{algorithmic}
\REQUIRE Integer $i$ belongs to the set of indices in the computational domain $D$.
\FOR{integration step s = 1 \TO 3 }
	\STATE{Compute boundary conditions for $\rho$ and $\mathbf{u}$}
	\FORALL {$i \in D$ \textbf{in parallel}} 
	\STATE{$(\nabla \cdot \mathbf{u})_{i} 
			\leftarrow \frac{\partial}{\partial x} u_{i} + \frac{\partial}{\partial y} v_{i} + \frac{\partial}	{\partial z} w_{i}$}
	\STATE {$\widetilde{\rho}_{i} 
			\leftarrow \alpha^{(s)}\widetilde{\rho}_{i} + \delta t\frac{\partial}{\partial t} \rho_{i}$}
	\STATE {$	\widetilde{\mathbf{u}}_{i} 
 			\leftarrow \alpha^{(s)} \widetilde{\mathbf{u}}_{i} 
                         + \delta_{t} \biggl[ -(\mathbf{u}_{i}\cdot\nabla)\mathbf{u}_{i}-c_s^2 \nabla \ln\rho_{i}
							+ \nu \bigg( \nabla^2\mathbf{u}_{i} + 2 \mathbf{S} \cdot \nabla \ln \rho_{i} \biggr)  \biggr]$}
	\STATE {$\rho_{i}^{\text{out}} 
			\leftarrow \rho_{i}+ \beta^{(s)}\widetilde{\rho}_{i}$}
	\STATE {$\mathbf{u}_{i}^{\text{out}} 
			\leftarrow \mathbf{u}_{i} + \beta^{(s)}\widetilde{\mathbf{u}}_{i}$}
	\ENDFOR 
	\STATE{Compute boundary conditions for $(\nabla \cdot \mathbf{u})$}
	\FORALL {$i \in D$ \textbf{in parallel}} 
	 \STATE{$\widetilde{\mathbf{u}}_{i} 
			\leftarrow \widetilde{\mathbf{u}}_{i} + \delta_{t} \frac{\nu}{3}\nabla(\nabla \cdot \mathbf{u})_{i}$}
	 \STATE{$\mathbf{u}_{i}^{\text{out}}
			 \leftarrow \mathbf{u}_{i}^{\text{out}} + \beta^{(s)} \delta_{t} \frac{\nu}{3}\nabla(\nabla \cdot \mathbf{u})_{i}$}
	\ENDFOR 
	 \STATE{$\rho \leftarrow \rho^{\text{out}}$}
	\STATE{$\mathbf{u} \leftarrow \mathbf{u}^{\text{out}}$}
\ENDFOR
\end{algorithmic}
\label{alg:19-point_method}
\end{algorithm}

\par

\begin{figure}
\centering
\includegraphics[width=0.44\linewidth]{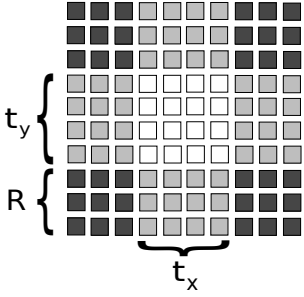}
\caption{The shared memory slab used in the 19-point 
method. The stencil computation is performed on the grid points in the center. Grey cells represent the 
grid points in the halo while the black areas are unused.}
\label{fig:19p_smem_slab} 
\end{figure}

We implemented the 19-point method in CUDA using the idea of 2.5D cache blocking \cite{mikki2009} \cite
{nguyen2010} to reduce the number of global memory transactions. In this approach, we set $\tau_z = 1$
and store a 2-dimensional slab of data into shared memory, shown in Fig. \ref{fig:19p_smem_slab}. For simplicity, we 
allocate shared memory for $(\tau_x + 2R) \times (\tau_y + 2R)$ grid points, where the four $R^2$-sized corners of 
the slab are unused. Since the shared memory slab contains only grid points in the
$xy$-plane, the rest of the stencil points required to solve the derivatives
with respect to the $z$-axis are defined as local variables, which are placed into registers by the compiler.
Similarly as in our 55-point method, each thread of a thread 
block then updates multiple grid points along the $z$-axis.
Cache blocking works in our implementation as follows.

\noindent\textbf{Initial step:}\\
(a) Assign a 2-dimensional block of grid points to a thread block.\\
(b) Fetch the data required for solving the the derivatives in the $xy$-plane
into shared memory\footnote{$(\tau_x + 2R) \times (\tau_y + 2R) - 4R^{2}$ grid points}
and the data required for solving the derivatives in the $z$-axis into the registers
of each thread\footnote{$2R$ grid points per thread}.\\
(c) Update the block of grid points using the data in shared memory and registers.\\

\noindent\textbf{Subsequent steps:} While a thread block has updated less than
$E_{z}$ blocks of grid points, do the following:\\
(d) Assign the next block of grid points in the $z$-axis to the thread block.\\
(e) Update the non-halo area\footnote{$(\tau_x \times \tau_y)$ grid points}
of the shared memory slab using data stored in the registers of the threads and update the halo\footnote{$(\tau_x + 2R) \times (\tau_y + 2R) - (\tau_x \times \tau_y) - 4R^{2}$ grid points} from global memory. For each thread, hold part of the data obtained in the previous steps in registers\footnote{$2R - 1$ grid points}, but update the local variable corresponding to the stencil point furthest along the $z$-axis from global memory.\\
(f) Update the assigned block of grid points using the data in shared memory and registers.

\par

Using the same notation as in Sect. \ref{sec:55-point_integration_method} and $\tau_z = 1$, the total number of grid points 
fetched from global memory by a thread block is

\begin{equation}
\Gamma_{19p} = \underbrace{\tau_{x}\tau_{y}E_{z} }_{\text{Computational area}} + \underbrace{2R(\tau_{x}\tau_{y} + \tau_{x}E_{z} + \tau_{y}E_{z})}_{\text{Halo area}} \ . 
\end{equation}

Additionally, by using Eq. \ref{eq:thread_blocks_per_grid} to solve the number of thread blocks $\gamma$ 
required to update the grid, we can now write the number of read-writes performed in the first and second pass as 
follows. During the first pass, we read-write to the same arrays as with the 55-point method in Eq. \eqref{eq:55p_rw} with the addition of
writing the divergence field to global memory. In the second pass, we read from one array including the halo
($\nabla \cdot\mathbf{u}^{(s)}$) and six partial result arrays ($\mathbf{\widetilde{u}}_{\text{partial}}$ and $\mathbf{u}_{\text{partial}}$),
and write to six arrays ($\mathbf{\widetilde{u}}$ and $\mathbf{u}$).

\begin{equation}
RW_{19p, \text{1st pass}} = \underbrace{4 \Gamma_{19p} \gamma + 4 N_{xyz}}_{\text{Reads}}  + \underbrace{9 N_{xyz} }_{\text{Writes}} 
\end{equation} 
and
\begin{equation}
RW_{19p, \text{2nd pass}} = \underbrace{\Gamma_{19p} \gamma + 6 N_{xyz}}_{\text{Reads}}  + \underbrace{6 N_{xyz} }_{\text{Writes}} \ .
\end{equation} 

\section{GPU performance}
\label{sec:results}

In this section, we present the results of our GPU implementations described in
Sections \ref{sec:55-point_integration_method} and \ref{sec:19-point_integration_method}. Additionally, we compare the
performance of our implementations with the Pencil Code \cite{pencil_code}, which
is a high-order finite-difference solver for compressible magnetohydrodynamic flows and is
developed to run efficiently on multi-CPU hardware. 
We strived to benchmark both our GPU implementations in a comparable way by using equally optimized
versions of both approaches. Additionally we compared the performance of our GPU implementations
with the Pencil Code using equally modern hardware sold at similar price points. 
The performance of our GPU implementations is not compared with any other GPU
finite-difference solver, because to our knowledge, no previous work has been done on simulating
compressible fluids on GPUs using sixth-order finite-differences.

\par

We generated the benchmarks for the implementations by running a test case, which simulated compressible
hydrodynamic flow by using sixth-order finite differences and third-order Runge-Kutta integration.
The benchmarks were run using single-precision floating-point numbers unless otherwise mentioned. Forcing was disabled in all performance tests. In order to get a fair comparison, we used grid sizes that are multiples of $12$ for generating 
the CPU results, since the workload in this case is divided more evenly on the $12$ cores
of an Intel Xeon E5-2690v3 processor. In contrast, the optimal grid sizes shared by both of our
GPU implementations are multiples of $32$, which we used to generate the GPU results. 
Diverging from our GPU implementations, Pencil Code uses a
2N-storage Runge-Kutta integration method which we described in Sect. \ref{sec:2N_RK_scheme}.
Therefore our 55-point method gives the same error as the Pencil Code, but our 19-point method 
does not. Additionally, we do not know whether the single-pass approach we used in our 55-point method would also
be suitable for CPUs, and how it would affect the performance if used within the Pencil Code. 

\par

We tested our GPU implementations on an NVIDIA Tesla K40t accelerator card, based on a single
875-MHz Kepler GK110BGL GPU (15 SMXs, 192 CUDA cores per SMX, 745 MHz base clock rate). The
on-device memory has a bus width of 384 bits and consists of a total of $12288$ MiB of GDDR5-3004
SDRAM (24 $\times$ 256MiB chips in clamshell mode), of which 11520 MiB is usable as
global memory. Tests were performed with ECC enabled. A compute node consists of two 2.6-GHz
Intel Xeon E5-2620-v2 Ivy Bridge CPUs (2.1 GHz base clock rate, 6 cores per CPU) with 32 GiB of
DDR3-1600 memory and two NVIDIA Tesla K40t accelerator cards, which are connected via a 16x PCI
Express 3.0 bus. We compiled the program with CUDA 6.5 and Intel 14.0.1 compiler (the only Intel compiler
supported in CUDA 6.5) for compute capability 3.5 invoking
\verb!nvcc! with flags \verb!-ccbin icc!, \verb!-O3! and \verb!-gencode arch=compute_35,code=sm_35!.

\par

The test case for Pencil Code was run on a compute node consisting of two 12-core
2.6-GHz Intel Xeon E5-2690v3 processors based on Haswell microarchitecture. Each core has 32 KiB
of L1 cache and 768 KiB of L2 cache while a 30-MiB L3 cache is shared by the cores of the CPU.  The
main memory of a node consists of 8 $\times$ 16 GiB DDR4-2133 DIMMs.
We used a revision of Pencil Code fetched on 2015-08-27
\footnote{Revision fdf3802edcd2d9036005534c94f288cd124a069a}. This build was compiled with a
Fortran 90 compiler for MPI programs invoked by the
\verb!mpif90! command using \verb!-O3!, \verb!-xCORE-AVX2!, \verb!-fma!, \verb!-funroll-all-loops!
and \verb!-implicitnone! flags. We used Intel compiler version 15.0.2 and Intel MPI library version 5.0.2.

\begin{figure*}
\centering
\begin{subfigure}{.5\textwidth}
  \centering
  \includegraphics[width=0.86\linewidth]{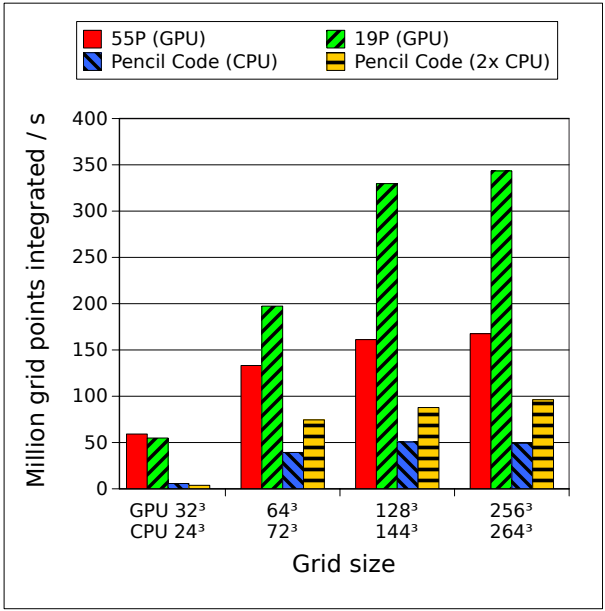}
  \caption{Performance comparison.}
  \label{fig:performance_comparison}
\end{subfigure}%
\begin{subfigure}{.5\textwidth}
  \centering
  \includegraphics[width=0.88\linewidth]{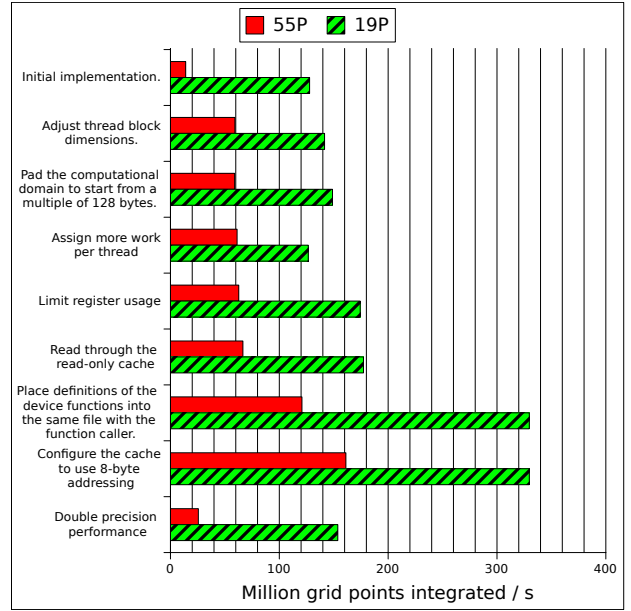}
  \caption{Optimization steps and double-precision results.}
  \label{fig:optimization_steps}
\end{subfigure}
\caption{(a) Performance comparison of the 19-point, 55-point methods and 
the CPU reference solver and (b) the optimizations performed upon the single-precision 
implementations and the double-precision results for our fastest solution,
using a $128^3$-sized grid.}
\label{fig:performance_results}
\end{figure*}

The performance comparison of our GPU implementations and the Pencil Code is
shown in Fig. \ref{fig:performance_comparison}. We achieved the rate of
$343$ million updates per second using the 19-point method, which was $2.0$ times faster than
integration with the 55-point method, which achieved the update rate of $168$ million updates per second. 
The Pencil Code achieved an update rate of $51$ million updates per second
using one CPU, while the update rate using two CPUs was $96$ million updates per second.
With $256^3$-sized grids and single-precision, we achieved the best performance for the 19-point method
by updating $16$ and $64$ grid points per thread in the first and second half of the algorithm,
respectively. With double-precision and the 55-point method, we had to decrease the size
of a thread block to $4 \times 4 \times 4$ threads in order to fit the required data
into shared memory. In this case, the best performance was achieved when a total of $32$ grid points
were updated by each thread.
With double-precision, we achieved the rate of $25$ and $154$ million elements integrated per
second with our 55-point and 19-point implementations, respectively.

\par

Fig. \ref{fig:optimization_steps} shows the optimizations performed upon our GPU implementations. 
Notably, the performance decreases during the optimization step, where more work is added per thread.
This is caused by overusing registers in the loop, which handles updating multiple grid points per thread, 
which in turn limits the occupancy of the integration kernel. As the next optimization step, we limited register usage
with the \verb!__launch_bounds__! intrinsic, which causes any additional local variable above the limit to spill into
L1 cache, but which in turn results in better performance because of the increased occupancy.
Without adding more work per thread, we could not increase the performance by limiting register usage
nor using the texture cache for reading. As the final optimization step, we moved the device functions used to compute
derivatives from separately compiled modules to the same source file with the integration kernel, which resulted in a large
boost in performance. The reason for this is, that when compiling a source, the
compiler cannot optimize any calls to functions in separately compiled units,
and must replace them with expensive calls which adhere to the application
binary interface
used in CUDA. With this step, we measured a $1.8\times$ and $1.9\times$ increase in performance of our 55-point and 19-point methods, respectively. 
We further improved the implementation of our 55-point method by setting the shared memory
configuration to use eight-byte addressing mode, which resulted in a speedup of $34\%$. 
This step reduced the bandwidth requirement for L1 and shared memory,
which was previously the limiting factor in the kernel. With the 19-point method,
we did not see any notable difference in performance.
The integration rate using double-precision is $2.1\times$ slower for the 19-point method and
$6.3\times$ slower for the 55-point method. Because of the increased shared memory requirements,
the latency issues in the 55-point method are accentuaded. In order to fit all the
required data into shared memory, smaller thread blocks have to be used, which reduces
both occupancy and the amount of data which can be reused. This causes the performance
to degrade more than $2 \times$ from the single-precision results. The 19-point
method is still bound by memory bandwidth, even though the increased shared memory requirements
limit the occupancy of the first half to $25\%$ or less regardless of the thread block
dimensions.

\par

Tables \ref{tab:kernel_results} and \ref{tab:db_kernel_results} show the
resource usage of our integration kernels. The 55-point method is limited by
low occupancy and is latency-bound, achieving the memory bandwidth of $72$ GB/s. The 19-point method achieved
$72\%$ and $79\%$ of the theoretical maximum bandwidth during the first and second pass, respectively. As the performance
of the kernels used in the 19-point method is mostly limited by the available bandwidth, the 19-point method is bandwidth-bound.
Occupancy of the 19-point method is limited by both shared memory and register usage. 

\begin{table}[htb]
\caption{The resource usage of the integration kernels using a $128^3$-sized grid, single-precision and a Tesla K40t GPU described in
Sect. \ref{sec:results}. 
Utilization of the load-store (LS), control flow (CF), arithmetic-logic unit (ALU) and texture function 
units (FUs) is measured in scale of 0 to 10. Throughput is measured in terms of million grid points 
updated per second. For reference, the peak effective bandwidth with ECC depends on the memory 
access pattern of the kernel and is expected to be roughly 80\% of the theoretical maximum 
bandwidth \cite[Sect. 8.2]{cuda_best_practices}. The theoretical maximum bandwidth of a Tesla K40t GPU is $288$ GB/s.}
\begin{center}
    {\small
    \begin{tabular}{ | l | l | l | l |}	
    \hline
     & 55P kernel & 19P kernel, & 19P kernel, \\
     &            & 1/2         & 2/2  \\
	\hline
	Thread block (TB)& & & \\dimensions & (8, 8, 8) & (32, 4, 1) & (32, 4, 1)\\
	\hline
	Grid point size& & & \\(bytes) & 16 & 16 & 4\\
	\hline
	TB shared memory& & & \\usage (bytes) & 43~904 & 6080 & 1520\\
	\hline
	32-bit registers per& & & \\CUDA thread  & 126 & 64 & 26\\
	\hline
	Grid points updated& & & \\per thread  & 8 & 8 & 1\\
	\hline
	Memory transactions& & & \\(million) & 9.3 & 8.2 & 4.3\\
	\hline
	L1/Shared memory & & & \\bandwidth (GB/s)  & 1374 &  1038 & 628\\
	\hline
	Device memory & & & \\bandwidth (GB/s)  & 72 &  203 & 227\\
	\hline
	Achieved occupancy & 25 \% & 49 \% & 93 \%\\
	\hline
	Occupancy limited by & \begin{tabular}{@{}c@{}}Cache and\\ registers\end{tabular} & \begin{tabular}{@{}c@{}}Cache and\\ registers\end{tabular} & n/a\\
	\hline
	Kernel bound by & Latency & Bandwidth & Bandwidth\\
	\hline
    \end{tabular}
    }
\end{center}
\label{tab:kernel_results}
\end{table}

\begin{table}[htb]
\caption{The resource usage of the integration kernels using a $128^3$-sized grid, double-precision and a Tesla K40t GPU.}
\begin{center}
    {\small
    \begin{tabular}{ | l | l | l | l |}	
    \hline
     & 55P kernel & 19P kernel, & 19P kernel, \\
     &            & 1/2         & 2/2  \\
	\hline
	Thread block (TB)& & & \\dimensions & (4, 4, 4) & (32, 4, 1) & (32, 4, 1)\\
	\hline
	Grid point size& & & \\(bytes) & 32 & 32 & 8\\
	\hline
	TB shared memory& & & \\usage (bytes) & 32~000 & 12~160 & 3040\\
	\hline
	32-bit registers per& & & \\CUDA thread  & 229 & 128 & 88\\
	\hline
	Grid points updated& & & \\per thread  & 32 & 32 & 64\\
	\hline
	Memory transactions& & & \\(million) & 24.6 & 16.7 & 9.1\\
	\hline
	L1/Shared memory & & & \\bandwidth (GB/s)  & 228 &  586 & 324\\
	\hline
	Device memory & & & \\bandwidth (GB/s)  & 29 &  204 & 191\\
	\hline
	Achieved occupancy & 3 \% & 25 \% & 29 \%\\
	\hline
	Occupancy limited by & Cache & \begin{tabular}{@{}c@{}}Cache and\\ registers\end{tabular} & Registers\\
	\hline
	Kernel bound by & Latency & Bandwidth & Bandwidth\\
	\hline
    \end{tabular}
    }
\end{center}
\label{tab:db_kernel_results}
\end{table}

\section{Physics tests} \label{sec:testcases}

We chose three simple test cases to validate the code physics-wise.
In the first test case we initialized a smooth sine-wave velocity profile
with varying wavenumber into the domain, and \textit{observed its decay due to
  diffusion}, for which an analytical solution can be derived. If we 
assume an initial condition where $\mathbf{u} = u_0 \sin(kx) \mathbf{\hat{e}}_y$ and a constant 
density, the velocity profile decays as a function of time into
\begin{equation}
\label{eq:dissipation}
\mathbf{u}(x,t) =  u_0 e^{-\nu t k^2} \sin(kx) \mathbf{\hat{e}}_y
\end{equation}
which can be compared with a snapshot calculated by the code. 

We performed the test both with the 19- and 55-point methods, and chose a high
wavenumber, $k=13$, to have a challenging test case for the
finite difference scheme.

\begin{figure}[h!]
\centering 
\includegraphics[width=0.5\textwidth]{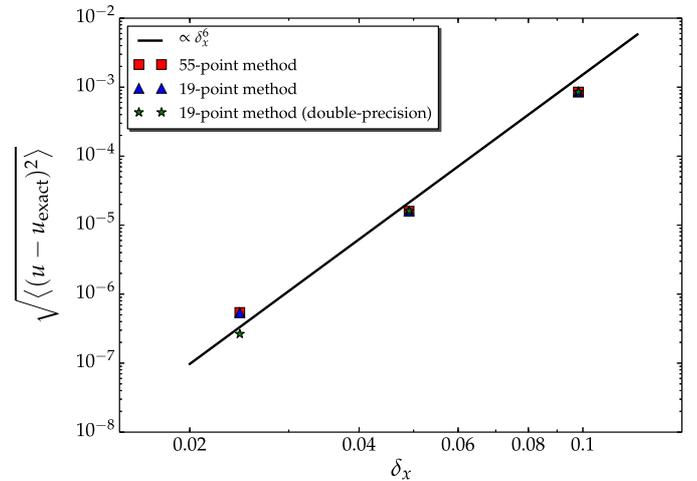} 
\caption{The root mean square of the point-by-point difference between a
snapshot and the analytical prediction (Eg. \ref{eq:dissipation}) at $t=1.5$
for both 55-point method and 19-point method with single- and double-precision. }
\label{fig:analytic_comparison}
\end{figure}

Our results are shown in Fig.~\ref{fig:analytic_comparison}, where
we plot the difference between the numerical and analytical solutions as a function
of $\delta_x$, and make a power-law fit to compare the trend with the expected one for
sixth-order finite differences, $\propto \delta_x^{6}$, as the resolution is
decreased. 

With both methods, the measured deviation from the analytical solution is
very close to the theoretical discretization error, the average change in
the deviation when halving the amount of grid points being $2^{5.7}$. 
Between the two highest resolution cases the wave is becoming too well
resolved, and the single grid point deviation is approaching the floating point
accuracy, and therefore the powerlaw starts showing hints of breaking down.
However, including double-precision, this breakdown is avoided at the highest resolution.
These results clearly show a satisfactory accuracy
of the finite-difference scheme with respect to the theoretical expectation.

\begin{figure}[h!]
\centering 
\includegraphics[width=0.5\textwidth]{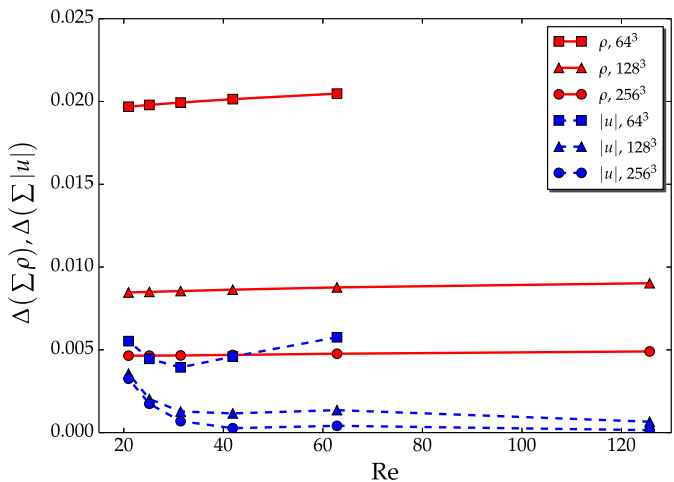} 
\caption{Relative directional errors of $\sum \rho$ and $\sum |u|$ at the end of the explosion tests.}
\label{fig:explosion}
\end{figure}

The second test case was a \textit{radial kinetic explosion}. The spherical 
symmetry of the test made it possible to check the symmetry of the
computational operation within the code, the rate-of-strain tensor, 
Eq. \eqref{eq:tensor}, in particular.
The kinetic energy was released in a small volume with a Gaussian
profile, directed  radially outward from the centre of the computational
domain. In  spherical coordinates, this initial condition was defined as 
\begin{equation}
\mathbf{u}(\mathbf{r}) = u_{\mathrm{ampl}}\exp\bigg( \frac{(r -
r_\mathrm{shell})^2}{2d_\mathrm{shell}^2} \bigg) \, \mathbf{\hat{e}}_r,
\end{equation}
where $r_\mathrm{shell}$ is the radius of the shell at the
peak, $u_{\mathrm{ampl}}$ the peak velocity and $d_\mathrm{shell}$ the
width parameter. The Gaussian profile helped to avoid numerical
instabilities due to discontinuous initial conditions. 
In a $64^3$, $128^3$- and $256^3$-sized grids we kept $u_{\mathrm{ampl}} = 1$, $d_\mathrm{shell} = 0.2$, $r_\mathrm{shell} = 0.8$, $\rho_0 = 1$ and varied 
$\mathrm{Re}$ (Fig. \ref{fig:explosion}) by changing the kinematic viscosity values. 
The tests were performed with both the 55-point and the 19-point method.

We measure the degree of spherical symmetry on a Cartesian grid
by comparing the values of the quantities along 7-directional axes 
with each other. Three were the 
$x$-, $y$-, $z$-axes and four diagonals which went from all corners to
their opposites. As we could not match the coordinates of Cartesian
and the diagonal axes point by point, we integrated the sum of
values of each axis $\big( \sum \rho \big)_i$
and $\big( \sum |u| \big)_i$. In the ideal spherically symmetric case,
the results 
for all axes should be the same, therefore we estimated the relative 
errors between axes by taking standard deviations 
$\Delta \big( \sum \rho \big)$ and $\Delta \big( \sum |u| \big)$. The
results are shown in Fig. \ref{fig:explosion} for the 19-point stencil method.
$\Delta \big( \sum |u| \big)$ is largest for the smallest Reynolds numbers
within a certain set with the same resolution. There is a strong decrease when
the Reynolds number is increased, and at the highest Reynolds numbers
investigated, the difference between the diagonal and off-diagonal elements
no longer changes for the highest resolution cases.
This is as expected, as the numerical solution is more likely to deviate
  from the analytic one at small Reynolds numbers where the viscous effects
  are stronger.
For the lowest resolution
case no convergence is seen, indicating that this test is too demanding to
be performed with that grid spacing.
There is a weak, but opposite, trend in $\Delta \big( \sum \rho \big)$.
This is likely to be due to the fact that the decrease of viscosity
enhances the effects of the non-conservative nature of the discretized
equations.
As the 55-point stencil method produces results that practically coincide
with the 19-point method, the results are shown in Fig.~\ref{fig:explosion}
only for the 19-point method.
This test shows that we can satisfactorily re-produce spherically symmetric
structures with the numerical scheme.

The third test case explored is a typical \textit{forced non-helical 
  turbulence setup}.
We switch on the external force term in the momentum equation, and use a
non-helical forcing function, that can be expressed as
\begin{small}
\begin{equation}
\mathbf{f}(\mathbf{x},t) = \mathrm{Re}\bigg\{ N \mathbf{f}_{\mathbf{k}(t)}
\exp\big(i \mathbf{k}(t) \cdot \mathbf{x} + i \phi(t) \big) \bigg\},
\label{eq:forcing}
\end{equation}
\end{small}
where 
\begin{small}
\begin{equation} 
N = f_0 c_s \sqrt{\frac{|\mathbf{k}| c_s}{\delta t}} \quad \mathrm{and} \quad  
\mathbf{f}_{\mathbf{k}(t)} = \frac{\mathbf{k} \times
\mathbf{e}}{\sqrt{|\mathbf{k}|^2 - (\mathbf{k} \cdot \mathbf{e})^2}}.
\end{equation}
\end{small}
Here $f_0$ scales the magnitude of the forcing and $\mathbf{e}$ is an arbitrary
unit vector perpendicular to forcing wave vector $\mathbf{k}$ \cite{Brandenburg2001ApJ}. An isotropic set of 
$\mathbf{k}$-vectors, with $k_\mathrm{min} \leq |\mathbf{k}| \leq k_\mathrm{max}$, is generated at the 
beginning of a computational run, from which they are picked randomly at every timestep. 

We denote the mean value of the set of vectors $|\mathbf{k}|$ as $k_f$, 
and chose sets with $k_f = 1.53$, $2.23$, $3.13$, $4.12$ and $10.0$.
With the default domain size $(2\pi, 2\pi, 2\pi)$ the box size corresponds
to $k = 1$.
We performed the forcing tests using the 19-point stencil method with
a $256^3$ grid, with $\nu = 1.4 \cdot 10^{-4}$, 
which was the lowest viscosity still stable with this resolution. 

\begin{figure}[htb]
 \centering
 \includegraphics[width=0.5\textwidth]{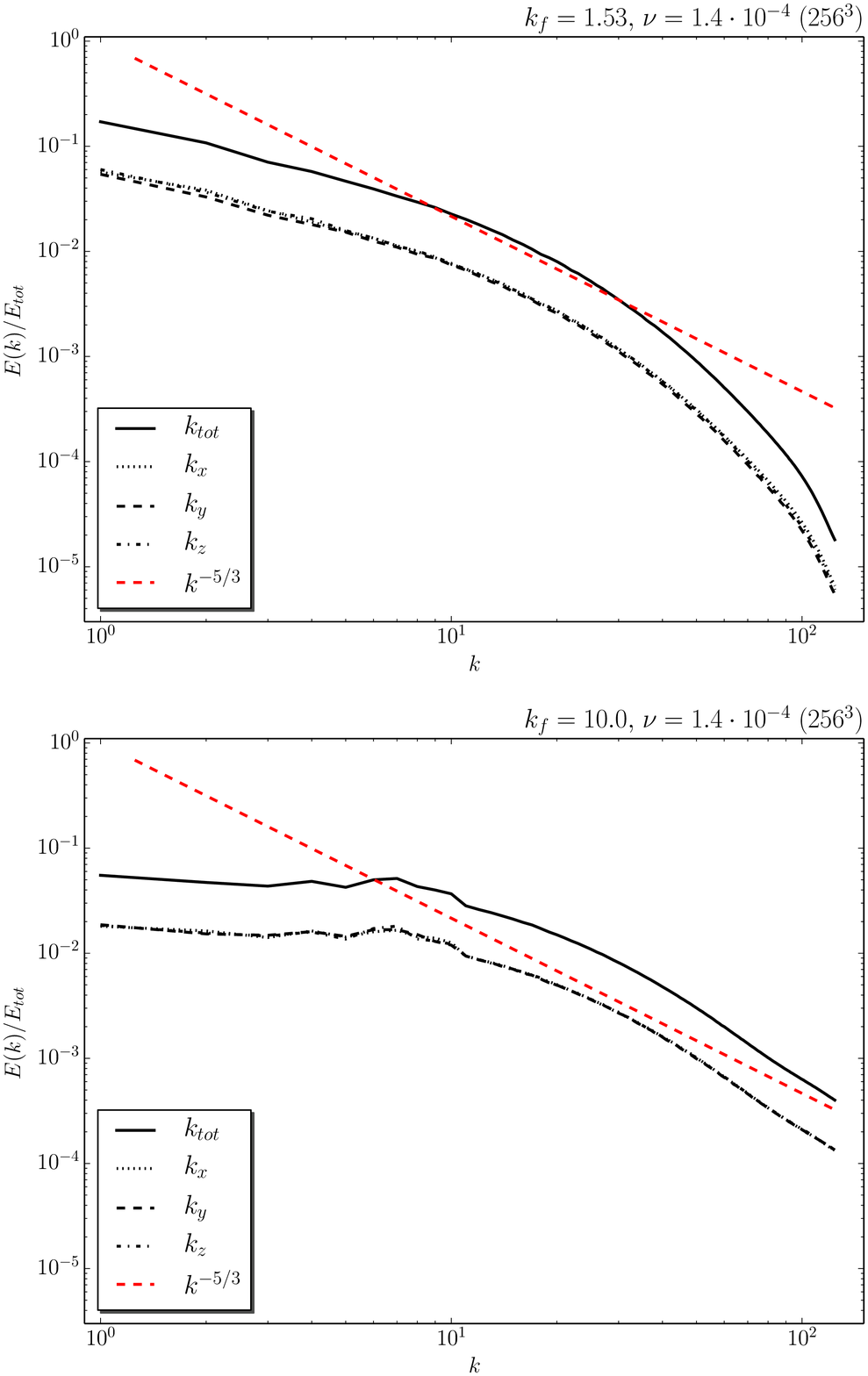} 
\caption{Normalized power spectra of saturated forced turbulence with $k_f \sim
1.53$ (\textit{top}) and $k_f \sim 10.0$ (\textit{bottom}).}
\label{fig:kolmogorov}
\end{figure}

The results of the forced turbulence test are shown as 
spectra in Fig. \ref{fig:kolmogorov}, which describe the
relative distribution of turbulent energy for different wavenumbers $k$. 
The peak of turbulent energy was situated at the wavenumber 
of the forcing $k_f$ and cascaded into 
smaller scales or higher wavenumbers.
In addition, the spectra show that turbulence behaves isotropically.  
Following the turbulence theory of Kolmogorov the kinetic energy is expected
to scale as $E(k) \propto \sim k^{-5/3}$.
However, the inertial range of the turbulence, where the energy distribution 
follows well the Kolmogorov theory, is quite limited at the Reynolds
numbers achievable with the current resolution and viscosity
scheme. To increase the effective Reynolds numbers, the implementation
of dynamic numerical diffusion schemes, such as shock and hyperviscosities
\cite{2001CauntKorpi} would be needed.
However, they are special features which are not relevant to the focus of 
this study.

\section{Discussion and Conclusions}
\label{sec:discussion}

We have implemented two methods based on 19-point and 55-point stencils for simulating compressible fluids on GPUs using 
third-order Runge-Kutta integration and sixth-order finite differences. Our
study is meant as a proof of concept for further developments with
  multi-GPU systems. The physical problem currently solved for is that of isothermal compressible hydrodynamics.

\par

We show that integration kernels, which operate on 55-point stencils and where each
stencil point holds several variables, three-dimensional cache blocking
is inefficient because of the limited amount of shared memory available on
current GPUs. While the performance of these kernels can be improved with
further optimizations, the latencies caused by arithmetic and memory operations
are difficult to hide with occupancy without introducing redundant
global memory fetches. Using instruction-level-parallelism to hide latencies is
similarly problematic, since large amounts of data are required to be cache
resident when processing several grid points at once.

We propose a reformulation of the problem, which can be used to improve the
performance of
the 55-point method by updating the simulation
grid in two passes using 19-point stencils. 
Our results show that the increased
occupancy and simpler memory access pattern in our 19-point method outweighs the
penalty of performing additional memory transactions.

Moreover, we show that the difference in accuracy of our 19-point and 55-point methods is minimal.
Other studies have shown that it is possible to achieve the bandwidth-bound
limit in stencil computation \cite{mikki2009} \cite{nguyen2010} \cite{zhang2012}, which
we also achieve with our 19-point implementation. 

We report the speedup of $1.7 \times$ and $3.6 \times$ between a 24 CPU
core node and a single GPU with the 55-point and the 19-point methods, respectively. 
Between a 12-core CPU and a GPU, the corresponding speedups are $3.3 \times$ and $6.8 \times$.
We consider this as a success as typical
supercomputer nodes have at least two GPUs, while the most recent
offerings of NVIDIA (DGX-1) have up to eight GPUs per node.
Although the final
outcome of the efforts depends crucially on how well the code scales
to multiple GPUs, the current performance is in our view clearly worth
the invested time and effort.

\par

We devote the rest of this section for explaining our design decisions and 
discussing the lessons learnt for future development. 
There are at least three ways to reduce the cache requirements in order to
improve latency-hiding in the 55-point method. 
First, smaller thread blocks can be 
used, which require fewer grid points to be cache-resident. However, since the size of the halo stays the same, this 
approach increases the ratio of memory transfers required for updating a grid point. In a $128^3$-sized grid, we
reached the update rate of only $85$ million points per second using $(32, 4, 1)$-sized thread blocks which updated $128$
points per thread and alternatively the update rate of $72$ million points per second with $(4, 4, 4)$-sized blocks updated $16$
points per thread.

\par

Second, part of the grid points can be stored into local memory while using 2.5-dimensional cache blocking. 
However, this in turn complicates the program when using 55-point stencils and introduces a large number of redundant 
fetches from the on-device memory, since only the grid points stored in registers within a warp can be shared with the 
shuffle instruction. For this reason, we did not explore this approach further.

\par

Finally, the integration kernel can be decomposed into a number of kernels, which use a smaller amount of resources 
than the initial kernel. As we show with our 19-point method, this optimization technique can be used to transform a 
latency-bound kernel into a number of kernels, which exhibit higher occupancy. Instead of decomposing the integration 
kernel of the 55-point method with respect to each dimension, we chose to solve the gradient of divergence separately, 
which in turn allowed us to use smaller stencils and 2.5-dimensional blocking efficiently, resulting in two
bandwidth-bound kernels. 

\par

However, the exact speedup gained from kernel decomposition is difficult to predict. If we assume that the time spent 
on computation in the initial kernel is negligible and that we achieve high enough occupancy with the decomposed 
kernels to be bandwidth-bound, we can approximate the potential maximum speedup of kernel decomposition with the 
following equations. 

\par

Let $BW_\mathrm{max}$ be the maximum effective bandwidth and $BW_\mathrm{initial}$
the bandwidth achieved with the latency-bound kernel. Additionally, let 
$M_\mathrm{initial}$ be the number of bytes transferred by the initial kernel and
$M_{i}$, where $i \in N$, be the number of bytes transferred by the $i$th kernel of $N$
decomposed kernels.

\par

With these definitions, we can write the time spent on transferring data
in the initial kernel and the time spent on transferring data in the decomposed kernels as

\begin{equation}
T_\mathrm{initial} = \frac{M_\mathrm{initial}}{BW_\mathrm{initial}} \ , \quad
T_\mathrm{decomposed} = \sum_{i=1}^{N} \frac{M_{i}}{BW_\mathrm{max}}\\
\end{equation}

Furthermore, we can now write the theoretical maximum speedup as

 \begin{align}
\frac{T_\mathrm{initial}}{T_\mathrm{decomposed}} &= \frac{M_\mathrm{initial}}{M_{1} + M_{2} + ... + M_{n}} \cdot \frac{BW_\mathrm{max}}{BW_\mathrm{initial}} \ .
\end{align}

\par

However, both our implementations perform more than the optimal amount of memory transaction to update 
the simulation grid. Since we store the grid as a structure of arrays and Kepler architecture GPUs service 
memory transactions from global memory and L2 cache in segments of $32$ bytes, memory bandwidth is wasted on 
transferring irrelevant bytes when fetching data to the halos on the left and right side of the shared memory 
block in the $x$-axis. For $R=3$ and using floating-point precision, one such memory transaction wastes 
bandwidth on transferring $24$ additional bytes. By using an array of structures to store the grid, we would 
expect a reduction of $25.0\%$ and $3.4\%$ in the number of transactions used to read data with our 55-point and 
the first half of the 19-point methods, respectively. 

\par

In addition, shared memory and memory bandwidth is wasted in our implementation of 55-point method, 
since we fetch $(\tau_x+2R) \times (\tau_y+2R) \times (\tau_z+2R)$ grid points into shared memory during the initial update even when
some of the data is used by only one thread, or none at all. We chose this approach for its simplicity and to avoid 
branching execution paths in the integration kernel. The optimal size for the shared memory block when
using 55-point stencils is hard to formulate, but with a numerical test we found that 
when $\tau_x = \tau_y = \tau_z = 8$ and $R=3$ we use approximately 
$34\%$ more shared memory with our 55-point method than the amount required for storing only 
those grid points, which are used by more than one thread. However, we do not know if using less shared memory would 
increase performance, as in addition to introducing branching to the algorithm, all grid 
points except the $R^3$-sized corners of the shared memory block fall on a memory segment, 
which is serviced from global memory or L2 cache with a single memory transaction in both
approaches regardless of whether the point is stored in shared memory.

\par

Additionally, our 19-point implementation is not compute-bound. Nguyen et. al \cite
{nguyen2010} suggest, that it is possible to be compute-bound in stencil computation by using temporal 
blocking, where a thread block updates a block of grid points over several timesteps and therefore larger amount of data in shared memory can be reused. However, they also state that temporal blocking requires a large cache to increase performance in 
stencil computation where $R$ and the size of a grid point is high \cite{nguyen2010}. In future work, we are planning to explore
this approach as well as storing the grid as an array of structures.

\par

Using single-precision and $256^3$-sized grids, we saw a $4\%$ improvement
in performance when integrating significantly more grid points per thread than in
our optimal solution for $128^3$-sized grids. A contributing factor was the following.
While increasing the workload per thread reduces
the number of thread blocks running on the GPU, with larger grids the 
number of thread blocks is sufficiently large to saturate the GPU with work
even when integrating a much larger number of grid points per thread.
Solving more grid points per thread increases the reuse factor, which in turn
reduces the relative number of global memory fetches needed to integrate a grid point.

\par

While the focus of this work was to accelerate integration using single-precision, 
we showed that the 19-point method scales well also to double-precision with minimal changes. 
With the 55-point method, we show that three-dimensional cache blocking becomes less efficient
if the data required to integrate a grid point is further increased.
Whether further decomposition could improve the performance over our
current best solution is a matter of future research.
Our tests without the rate-of-strain tensor, $\mathbf{S}$, indicate that the performance of the
first half of the 19-point method can potentially be improved by increasing occupancy of the kernel
by reducing the requirements for shared memory and registers, which in turn pushes the bandwidth achieved
closer to the hardware maximum. However, in order to avoid introducing a large number of additional
global memory fetches, as much of the data used for integrating a grid point should be retained
in caches until the data is no longer needed.

\par

Also one possible way to reduce shared memory requirements is to decouple the computation with the
velocity components and solve the results with respect to one axis at a time. Instead of
defining four arrays in shared memory as in the current kernel implementation, we could
limit their number to two. In order to improve latency hiding, we could also interleave
compute and memory instructions by using a technique called Ping-Pong buffering, where
the data is being processed in the 'Ping' buffer while new data is being fetched into the 'Pong' buffer.
However, a thorough investigation is required to determine whether these approaches
can be used to increase the performance of our solver.

\par

We are also planning to extend our implementations to support the full MHD
equations. These equations would require additional arrays to be stored in
global memory, such as the thermal energy and magnetic fields of the fluid. 
Since the addition of these new arrays increase the size of a grid point,
special care must be taken in order to avoid using large amounts of shared
memory. With a single-pass approach, such as our 55-point method, we would
expect the occupancy to degrade further if more data is required to
be cache-resident during an integration step. Thus, we would expect kernel 
decomposition to provide speedups also with the full MHD equations, 
since less data need to be stored in caches at a time in a multi-pass approach.

\par

On the other hand, the latest GPU architectures geared towards scientific
computing employ larger caches. For example, a Tesla K80 contains
$114 688$ bytes of shared memory. In our current 55-point implementation,
this would allow two $(8, 8, 8)$-sized thread blocks to be multithreaded on a SIMD processor
instead of only one on a Tesla K40t. However, we did not have access to a
Tesla K80 and do not know if the increased occupancy is high enough to
hide the latencies in our 55-point method and to outperform our 19-point
method.

\par

Additionally, a natural follow-up to our implementations is
to extend them to work with multi-node GPUs. We expect inter-node communication
to become the bottleneck, because of the comparably slow PCIe bus and communication
required for solving boundary conditions. However, an important benefit of
a multi-node implementation is that it would allow us to handle larger grids
than the current maximum of $512^3$ on a Tesla K40t.

\appendix


\section{}
The reformulation of the Navier-Stokes equations for the 19-point method is done as follows. For the intermediate result holds that

\begin{align*}
\widetilde{\mathbf{u}}^{(s+1)} 
&= \alpha^{(s)} \widetilde{\mathbf{u}}^{(s)} + \delta_{t} \frac{\partial}{\partial t} \mathbf{u}^{(s)}
\\
&= \alpha^{(s)} \widetilde{\mathbf{u}}^{(s)} 
+ \delta_{t} \biggl[ -({\mathbf{u}}^{(s)}\cdot\nabla){\mathbf{u}}^{(s)} -c_s^2 \nabla \ln\rho^{(s)} \\
&+ \nu \bigg( \nabla^2 \mathbf{u}^{(s)}  + \frac{1}{3}\nabla(\nabla \cdot \mathbf{u}^{(s)}) + 2 \mathbf{S} \cdot \nabla \ln \rho^{(s)} \biggr) \biggr]
\\
&= \alpha^{(s)} \widetilde{\mathbf{u}}^{(s)} 
+ \delta_{t} \biggl[ -({\mathbf{u}}^{(s)}\cdot\nabla){\mathbf{u}}^{(s)} -c_s^2 \nabla \ln\rho^{(s)} \\ 
&+ \nu \bigg( \nabla^2\mathbf{u}^{(s)} + 2 \mathbf{S} \cdot \nabla \ln \rho^{(s)} \biggr)  \biggr]
+ \delta_{t} \frac{\nu}{3}\nabla(\nabla \cdot \mathbf{u}^{(s)}) \ .
\end{align*}

When we set
\begin{align*}
\widetilde{\mathbf{u}}_{partial}^{(s+1)} 
&= \alpha^{(s)} \widetilde{\mathbf{u}}^{(s)} 
+ \delta_{t} \biggl[ -({\mathbf{u}}^{(s)}\cdot\nabla){\mathbf{u}}^{(s)} -c_s^2 \nabla \ln\rho^{(s)} \\
& + \nu \bigg( \nabla^2\mathbf{u}^{(s)} + 2 \mathbf{S} \cdot \nabla \ln \rho^{(s)} \biggr)  \biggr] \ ,
\end{align*}

it follows that
\begin{equation*}
\widetilde{\mathbf{u}}^{(s+1)} = 
\widetilde{\mathbf{u}}_{partial}^{(s+1)} + \delta_{t} \frac{\nu}{3}\nabla(\nabla \cdot \mathbf{u}^{(s)}) \ .
\end{equation*}

Likewise for the final result

\begin{align*}
\mathbf{u}^{(s+1)} 
&= \mathbf{u}^{(s)} + \beta^{(s)} \widetilde{\mathbf{u}}^{(s+1)}
\\
&= \mathbf{u}^{(s)} + \beta^{(s)} \biggl( \widetilde{\mathbf{u}}_{partial}^{(s+1)} + \delta_{t} \frac{\nu}{3}\nabla(\nabla \cdot \mathbf{u}^{(s)}) \biggr)
\\
&= \mathbf{u}^{(s)} + \beta^{(s)} \widetilde{\mathbf{u}}_{partial}^{(s+1)} + \beta^{(s)} \delta_{t} \frac{\nu}{3}\nabla(\nabla \cdot \mathbf{u}^{(s)}) \ .
\end{align*}

When
\begin{equation*}
\mathbf{u}_{partial}^{(s+1)}
= \mathbf{u}^{(s)} + \beta^{(s)} \widetilde{\mathbf{u}}_{partial}^{(s+1)} \ ,
\end{equation*}

it follows that
\begin{equation*}
\mathbf{u}^{(s+1)} 
= \mathbf{u}_{partial}^{(s+1)} + \beta^{(s)} \delta_{t} \frac{\nu}{3}\nabla(\nabla \cdot \mathbf{u}^{(s)}) \ .
\end{equation*}

\section*{Acknowledgements}

M.V. thanks financial support from Jenny and Antti Wihuri Foundation and
Finnish Cultural Foundation grants.
J.P. thanks Aalto University for financial support.
Financial support from the Academy of Finland through the ReSoLVE
Centre of Excellence (JP, MJK, PJK; grant No. 272157) and the
University of Helsinki research project `Active Suns' (MV, MJK, PJK)
is acknowledged.
We thank Dr. Matthias Rheinhardt for his
helpful comments on this paper and Prof.~Petteri Kaski for general guidance. 
We acknowledge CSC -- IT Center for Science Ltd., who are administered by the
Finnish Ministry of Education, for the allocation of computational resources.
This research has made use of NASA’s Astrophysics Data System.

  \bibliographystyle{elsarticle-num} 
  \bibliography{references}





\end{document}